\documentclass[pdflatex,sn-mathphys-ay]{sn-jnl}% Math and Physical Sciences Author Year Reference Style

\usepackage{graphicx}%
\usepackage{multirow}%
\usepackage{amsmath,amssymb,amsfonts}%
\usepackage{amsthm}%
\usepackage{mathrsfs}%
\usepackage[title]{appendix}%
\usepackage{xcolor}%
\usepackage{textcomp}%
\usepackage{manyfoot}%
\usepackage{booktabs}%
\usepackage{algorithm}%
\usepackage{algorithmicx}%
\usepackage{algpseudocode}%
\usepackage{listings}%
\usepackage{natbib}

\newcommand{\bm}[1]{ \mbox{\boldmath $#1$} }
\theoremstyle{thmstyleone}%
\newtheorem{theorem}{Theorem}%  meant for continuous numbers
\theoremstyle{thmstyletwo}%

\theoremstyle{thmstylethree}%

\begin{document}

\title[Bayesian multivariate models for bounded
circular data]{Bayesian multivariate models for bounded
directional data}


\author[1]{\fnm{Joel} \sur{Montesinos--V\'azquez}}\email{joel.monva@gmail.com}

\author*[2]{\fnm{Gabriel} \sur{N\'u\~nez--Antonio}}\email{gabo.nunezantonio@gmail.com}
\equalcont{These authors contributed equally to this work.}

\affil[1]{\orgname{Universidad Aut\'onoma Metropolitana--Unidad Iztapalapa}, \orgaddress{\street{Av. San Rafael Atlixco 186}, \city{Alc. Iztapalapa}, \postcode{09340}, \state{Mexico City}, \country{Mexico}}}

\affil*[2]{\orgdiv{Department of Mathematics}, \orgname{Universidad Aut\'onoma Metropolitana--Unidad Iztapalapa}, \orgaddress{\street{Av. San Rafael Atlixco 186}, \city{Alc. Iztapalapa}, \postcode{09340}, \state{Mexico City}, \country{Mexico}}}

\abstract{In some areas of knowledge there are data representing directions restricted to a specific range of values. Consequently, it is useful to have models for describing variables defined in subsets of the \textit{k}-dimensional unit sphere. This need has led to the development of models such as the multivariate projected Gamma distribution. However, the proposal of multivariate models whose marginal variables are defined only in sections of the unit circle and with a flexible dependency structure is limited. In this work, we propose constructing multivariate models where each marginal variable is a circular variable defined only in the first quadrant of the unit circle. Our approach is based on  the concept of copula functions. The inferences for the proposed models rely on generating samples of the posterior joint density of all parameters involved in the models. This is achieved by applying a conditional approach that allows inferences to be made using a two-stage sampling. The proposed methodology is illustrated with both simulated and real data.}

\keywords{Copulas, Multivariate Circular Data, the Projected Gamma distribution, Conditional Sampling}

\maketitle

\section{Introduction}\label{sec1}

The study of directional variables arises from the need to account for the angular or periodic nature of special observations. Some applications of this type of data can be found in the analysis of wind directions, the orientation of migratory birds, and the record of mammals sighting throughout the day in their natural habitat, among others. This necessity has led to the development of probability models to describe this type of data, where the natural sample space is the $k$-dimensional unit sphere, $\mathbb{S}^k$. These variables representing directions can be considered as points on $\mathbb{S}^k$ or  unit vectors $\bm{x}$ in $\mathbb{R}^{k+1}$. In case $k=1$, $\mathbb{S}^k$ turns out to be the unit circle and the variables $\bm{x}$ can also be described by an angle $\theta$ defined on $[0,2\pi)$. In this case, the variable $\theta$ is said to be a circular variable. In case $k=2$, the directional variables are known as spherical variables. For a survey from the frequentist perspective, the reader is referred to \cite{fisher1995}, \cite{mardiajupp2000}, and more recently, to \cite{pewseygp2021}.

On the other hand, in some fields of knowledge there are studies involving circular data, such as shooting angles, tilt angles, or angles related to orthopedic gait, among others (\cite{jiaetal2024}, \cite{yunus2020}). Even though the above data can be considered circular variables, these are defined only in a section of the unit circle. In such applications, it is natural to assume that the angles are restricted to the positive orthant of the unit circle, that is, $\theta \in [0,\frac{\pi}{2}]$. In contrast to the attention that circular variables have received in recent years, bounded circular variables  or defined only on a subset of $\mathbb{S}$ seem to have been overlooked until now. Some notable exceptions are  \cite{nunezgeneyro2021}, \cite{guardiola2020} and \cite{harris2024}.

An additional problem arises when working with multivariate circular random variables. That means, random vectors on $\mathbb{S}\times\mathbb{S}\times\ldots\times\mathbb{S}$. The first models developed on the unit torus $\mathbb{S}\times\mathbb{S}$ were the von Mises bivariate distribution \cite{mardia1975} and the wrapped bivariate normal distribution \cite{johnsonwehrly1977}. However, the difficulty in interpreting the parameters and the limited possibility of carrying out efficient inferences, had led to research into alternative models. A more flexible proposal given by \cite{wehrlyjohnson1980} was to consider bivariate densities defined by
\begin{equation}\label{whrlyjohn}
f(\theta_1,\theta_2)=2\pi\cdot g(2\pi(F_1(\theta_1)\pm F_2(\theta_2)))f_1(\theta_1)f_2(\theta_2),
\end{equation}
where $f_1$ and $f_2$ are circular densities, and $F_1$ and $F_2$ are the corresponding distribution functions. Here $g(\cdot)$ is a circular density too. Since the functions $f_1$ and $f_2$ turn out to be the marginal densities, the model (\ref{whrlyjohn}) is more flexible. However, since the function $g$ has to be a density on $\mathbb{S}$, it restricts the options of possible models. Following these ideas,\citep{shiehj2005} define a density on the unit torus whose marginal densities are univariate von Mises distributions. \citep{singh2002} propose to replace linear and quadratic terms in bivariate normal densities with their circular analogues. In the same venue, another interesting alternative to build multivariate models is the use of copula functions. In this context, the connection between the joint distribution and the corresponding marginal distributions is naturally recognized as
\begin{equation}\label{copula}
F(x,y)=C(F(x),F(y)),
\end{equation}
where $C$ is a copula function. \cite{shiehj2005} point out that there is a remarkable relationship between (\ref{copula}) and (\ref{whrlyjohn}). Although this framework has been mentioned by \cite{leyverd2017} and studied in some depth by \cite{garciap2013}, \cite{jonespewseykato2015} and \cite{katopewseyjones2018}, a more profound and general treatment for the use of copulas in directional multivariate models is still lacking. Additionally, the proposed models have marginal densities defined over the entire circle $\mathbb{S}$, omitting cases where the sample space is only a section of $\mathbb{S}$.

In this work we propose a methodology for building probability models defined on $\mathbb{H}=\mathbb{S}^+\times\mathbb{S}^+\times\ldots\times\mathbb{S}^+$, where $\mathbb{S}^+$ is the first quadrant of the unit circle. This is done using copula theory. Particularly, we introduce the \textit{$w$-coupled Projected Gamma} model. We implement a full Bayesian analysis for the proposed models  based on a conditional sampling approach. The procedure is shown through several examples.

The rest of this paper proceeds as follows. In Section \ref{sec2}, we briefly review some concepts of the copula theory, emphasizing implicit copulas. In Section \ref{sec3}, we present our proposal for building multivariate bounded-circular data and the way to perform Bayesian inferences. In addition, we describe two special cases of the $w$-coupled Projected Gamma model, namely, the Gaussian-coupled Projected Gamma model and the $t$-coupled Projected Gamma model. Section \ref{sec4} illustrates our proposal with some simulated examples and a real dataset on angles into an orthopedic context. Finally, Section \ref{sec5} provides some concluding remarks.

\section{Copulas}\label{sec2}

Although other procedures can be found in the literature, copula theory provides a natural way to construct multivariate distributions. This framework is based on linking univariate marginal distributions with a joint distribution by means of a copula function, as established by the Sklar's theorem. Next, we briefly introduce some key concepts about copulas. For a survey  the reader is referred to \cite{nelsen2006} and \cite{joe2017} and the references therein.

A copula in $\mathbb{R}^m$ is a function  $C:[0,1]^m\rightarrow [0,1]$ that satisfies:
\begin{enumerate}
	\item $C(u_1,u_2,\ldots,u_m)=0$, if $u_j=0$ for some $1\leq j\leq m$.
	\item $C(1,1,\ldots,u_i,\ldots,1)=u_i$. 
	\item For $\{a_1,\ldots,a_m\}$ and $\{b_1,\ldots,b_m\}$ such that $0\leq a_i\leq b_i\leq 1$:
	\begin{equation*}
		\sum_{i_1=1}^2\ldots \sum_{i_m=1}^2 (-1)^{i_1+\ldots+i_m} C(u_{1,i_1},\ldots,u_{m,i_m})\geq 0,
	\end{equation*}
	where $u_{j,1}=a_j$ and $u_{j,2}=b_j$ for $1\leq j\leq m$.
\end{enumerate}

The following result is a fundamental theorem for understanding the importance of copulas in the context of multivariate distribution functions.

\begin{theorem}[Sklar's Theorem]\label{teosklar}
If $F(X_1,\ldots,X_m)$ is a $m$-variate distribution function with univariate marginal distributions  $F_1(X_1),\ldots,F_m(X_m)$, then there is a copula $C$ such that  
\begin{equation*}
F(X_1,\ldots,X_m)=C(F_1(X_1),\ldots,F_m(X_m)).
\end{equation*}
Reciprocally, if $C$ is a $m$-variate copula and   $F_1(X_1),\ldots F_m(X_m)$ are univariate distribution
functions, then
\begin{equation}\label{eq1}
C(F_1(X_1),\ldots,F_m(X_m))
\end{equation}
is a $m$-variate distribution function.
\end{theorem}

The previous result establishes the relationship between a joint distribution and the corresponding univariate marginals, as well as a way to construct multivariate distributions.  Alternatively, if we define a multivariate distribution function as in the Eq. (\ref{eq1}), then the corresponding probability density function (pdf) is given by
\begin{equation*}
f(\bm{x})=c((F_1(x_1),\ldots,F_m(x_m))  f_1(x_1) \cdots f_m(x_m),
\end{equation*}
$\bm{x}=(x_1,\ldots,x_m)$, $f_1,\ldots,f_m$ are the density functions related to $F_1,\ldots,F_m$, and $c$ is the derivative of $C$ given by
\begin{equation*}
	c(u_1,\ldots,u_m)=\frac{\partial^m}{\partial u_1\ldots \partial u_m} C(u_1,\ldots,u_m).
\end{equation*}
The function  $c(\cdot)$ is called  {\it the copula density} $c$.

In the literature, several parametric families have been employed as copula functions. The most used are Archimedean copulas (Frank, Gumbel, Clayton) or implicit copulas ($t$, Gaussian). Other families include empirical copulas or vine copulas. In this work we  focus on implicit copulas.

\subsection{Implicit Copulas}

From the Sklar's Theorem (Theorem \ref{teosklar}), if $H$ is a $m$-variate distribution function, whose marginal distribution functions $H_1,\ldots, H_m $ are continuous, then there is a copula $C_H$ such that
\begin{equation}\label{eq3}
H(x_1,\ldots,x_m)=C_H(H_1(x_1),\ldots,H_m(x_m)).
\end{equation}

If the distribution functions $H_j$ are  invertible, then $H_j^{-1}(u)$ is defined for all $u\in [0,1]$. Thus, from Eq.  (\ref{eq3}), we obtain that

\begin{equation}\label{eq4}
\begin{split}
H&(H_1^{-1}(u_1),\ldots,H_m^{-1}(u_m))\\ &=C_H(H_1(H_1^{-1}(u_1)),\ldots,H_m(H_m^{-1}(u_m)))\\
&=C_H(u_1,\ldots,u_m).
\end{split}
\end{equation}

The Eq. (\ref{eq4}) shows how to construct a copula from a multivariate distribution $H$ and its marginal distribution functions $H_j $. This process, defined by the equation (\ref{eq4}), is based on the well-known {\it inversion method}, and the copula generated by this method is called an {\it implicit copula}. See, for example, \cite{nelsen2006} and \cite{smith2023}.

Two families of implicit copulas, which use elliptic distribution functions as particular cases for the $H$ distribution, are the Gaussian copula and the $t$ copula. Next, we specialize the discussion for these copulas.

{\bf The $t$ copula}.  In this case, $H=T_m(\cdot|\nu,\bm R)$ where $T_m(\cdot|\nu,\bm R)$ is the distribution function of a $m$-variate $t$-Student variable with $\nu$ degrees of freedom, correlation matrix $\bm R$ and mean vector the vector $\bm 0$.  From Eq. (\ref{eq4}) the $t$ copula with parameters $\nu$ and $\bm R$  is given by
\begin{equation}\label{copt}
C_T(\bm{u}|\nu,\bm R)=T_m(T^{-1}(u_1|\nu),\ldots,T^{-1}(u_m|\nu)\ | \ \nu,\bm{R}),
\end{equation}
where $\bm{u}=(u_1,\ldots,u_m)$ and $T(\cdot|\nu)$ is the distribution function of a univariate $t$-Student variable with $\nu$ degrees of freedom. The corresponding copula density is given by
\begin{equation}\label{dcopt}
c_T(\bm{u}|\nu,\bm{R})=\dfrac{t_m(T^{-1}(u_1|\nu),\ldots,T^{-1}(u_m|\nu)|\nu,\bm R)}{t(T^{-1}(u_1|\nu)|\nu)\cdots t(T^{-1}(u_m|\nu)|\nu)},
\end{equation}
where $t_m(\cdot|\nu,\bm{R})$ is the density function related to the distribution  $T_m(\cdot|\nu,\bm R)$, and $t(\cdot|\nu)$ is the corresponding density function of the distribution $T(\cdot|\nu)$.

{\bf The Gaussian copula.} Let $\Phi_m(\cdot|\bm R)$ and $\phi_m(\cdot|\bm R)$ be the distribution and density functions of a $m$-variate normal variable with correlation matrix $\bm R$ and mean vector the vector  $\boldsymbol{0}$, respectively. If  $\Phi$  and $\phi$  are the univariate standard normal density and distribution functions, respectively, the Gaussian copula is given by
\begin{equation}\label{copg}
C_G(\bm{u}|\bm R)=\Phi_m(\Phi^{-1}(u_1),\ldots,\Phi^{-1}(u_m)|\bm R),
\end{equation}
and the corresponding density copula is defined as
\begin{equation}\label{dcopg}
c_G(\bm{u}|\bm{R})=\dfrac{\phi_m(\Phi^{-1}(u_1),\ldots,\Phi^{-1}(u_m)|\bm{R})}{\phi((\Phi^{-1}(u_1))\cdots \phi(\Phi^{-1}(u_m))}.
\end{equation}

\section{The multivariate proposal}\label{sec3}

In this section, we propose a method to build joint models for bounded directional variables. Specifically, multivariate models in which the marginal variables are circular and defined only in the first orthant of the unit circle. Our approach is based on the construction of a copula model in which the marginal densities are Projected Gamma distributions. Firstly, we provide a brief introduction to the Projected Gamma model.

\subsection{The projected gamma model}

While circular data models have received considerable attention in recent decades, models for bounded circular variables have received little attention.
In this work we assume that the circular variables follow a projected Gamma distribution, which is a model defined only on the positive orthant. This model was introduced by \cite{nunezgeneyro2021}. They show how to carry out Bayesian inference based on a Gibbs sampling scheme after the introduction of suitably chosen latent variables. In the circular case, a variable $\theta$ is said to have a projected Gamma distribution, if its density function is given by
\begin{equation}\label{eq5}
PG(\theta|\boldsymbol\alpha,\beta)=\dfrac{\beta^{\alpha_2}\cos(\theta)^{\alpha_1-1}\sin(\theta)^{\alpha_2-1}}{B(\alpha_1,\alpha_2)(\cos(\theta)+\beta\sin(\theta))^{\alpha_1+\alpha_1}},
\end{equation}
where $\theta\in[0,\frac{\pi}{2}]$, $\boldsymbol\alpha=(\alpha_1,\alpha_2)$ and $\alpha_1,\alpha_2,\beta>0$.

\subsection{Multivariate models for bounded circular variables}

The proposal to construct multivariate densities is based on Sklar's theorem. Let $F_{PG}(\theta|\bm\alpha, \beta)$ and $
PG(\theta|\bm\alpha, \beta)$ denote the distribution and density functions of a projected Gamma variable with  parameters $\bm\alpha$ and $\beta$, respectively. If $C_w$ is a $m$-variate copula with parameter vector  $\bm{\omega_C}$, then
\begin{equation*}
C_w(F_{PG}(\theta_1|\bm\alpha_1, \beta_1),\ldots,F_{PG}(\theta_m|\bm\alpha_m, \beta_m) \ | \ \bm{\omega}_C)
\end{equation*}
is a distribution function of a $m$-variate vector $\bm\theta=(\theta_1,\ldots,\theta_m)$ whose marginal distributions are the projected Gamma distributions $F_{PG}(\theta_j|\bm\alpha_j,\beta_j)$, where $\bm\alpha_j=(\alpha_{j1},\alpha_{j2})$, $j=1,\ldots,m$. 

In addition, the corresponding density function is given by
\begin{equation}\label{eqPGwC}
\begin{split}
	&C_wPG(\bm\theta|\bm\alpha_1,\beta_1,\ldots,\bm\alpha_m, \beta_m,\bm\omega_C)  = \\
&c_w(F_{PG}(\theta_1|\bm\alpha_1, \beta_1),\ldots,F_{PG}(\theta_m|\bm\alpha_m, \beta_m) \ | \ \bm{\omega}_C)\\ 
&\times PG(\theta_1|\bm\alpha_1, \beta_1)\cdots PG(\theta_m|\bm\alpha_m, \beta_m).
\end{split}
\end{equation}
This model will be called a \textit{$w$-coupled Projected Gamma} model and denoted as $C_wPG$. It should be noted this model is a multivariate model,  in which the marginal densities $PG$ are defined on the arc $[0,\frac{\pi}{2}]$.

\subsection{Bayesian inference}

In order to have a more general exposition, in this section we will consider that the distribution function of the $w$-coupled model Projected Gamma can be written as
\begin{equation}\label{eq6a}
F(\bm{\theta}|\bm{\omega}_\Theta,\bm{\omega}_C)=C_w(F_1(\theta_1|\bm{\omega}_1),\ldots,F_m(\theta_m|\bm{\omega}_m)|\bm{\omega}_C),
\end{equation}
where $\bm{\omega}_\Theta=(\bm{\omega}_1,\dots,\bm{\omega}_m)$ represents the parameter vector of all marginal densities, and $\bm{\omega}_C$ is the parameter vector of the corresponding copula function. Here $F_j$ and $f_j$ denote the distribution and density functions, respectively, of the circular variable $\theta_j,$ $\forall \ j=1,\dots,m$.

Thus, the corresponding density function can be expressed as
\begin{equation*}
\begin{split}
f(\bm{\theta}|\bm{\omega}_\Theta,\bm{\omega}_C)&=c_w(F_1(\theta_1|\bm{\omega}_1),\ldots,F_m(\theta_m|\bm{\omega}_m)|\bm{\omega}_C)\\
& \times f_1(\theta_1|\bm{\omega}_1)\cdots f_m(\theta_m|\bm{\omega}_m).
\end{split}
\end{equation*}
When $\theta_j$ has a projected Gamma density, then  $f_j$ is a $PG(\theta_j|\bm{\omega}_j)$ model (Eq. \ref{eq5})  where $\bm{\omega}_j=(\bm{\alpha}_j,\beta)$ and   $F_j$ is the corresponding distribution function $F_{PG}$.

If $\bm{D}_n=\{\bm\theta_1,\ldots,\bm\theta_n\}$ is a sample of size  $n$ from a $C_wPG(\bm{\theta}|\bm{\omega}_{\Theta},\bm{\omega}_C)$ model (Eq. \ref{eqPGwC}), where $\bm\theta_i=(\theta_{i1},\ldots,\theta_{im})$, the posterior density of $(\bm{\omega}_\Theta,\bm{\omega_C})$ is given by

\begin{equation}\label{eq6}
\begin{split}
p&(\bm{\omega}_\Theta,\bm{\omega}_C|\bm{D}_n)\\
&=\frac{p(\bm{D}_n|\bm{\omega}_\Theta,\bm{\omega}_C)\ p(\bm{\omega}_\Theta,\bm{\omega}_C)}{\int \int p(\bm{D}_n|\bm{\omega}_\Theta,\bm{\omega}_C)p(\bm{\omega}_\Theta,\bm{\omega}_C)\  d\bm{\omega}_\Theta\ d\bm{\omega}_C },
\end{split}
\end{equation}
where 
\begin{equation*}
p(\bm{D}_n|\bm{\omega}_\Theta,\bm{\omega}_C)=\prod_{i=1}^n C_wPG(\theta_{i1},\ldots,\theta_{im}|\bm{\omega}_\Theta,\bm{\omega}_C)
\end{equation*}
and $p(\bm{\omega}_\Theta,\bm{\omega}_C)$ is the prior distribution of  the parameter vector $(\bm{\omega}_\Theta,\omega_C)$.

If the density $p(\bm{\omega}_\Theta,\bm{\omega}_C | \bm{D}_n)$ is not easy to analyze, an option is to use resampling techniques to make inferences for the parameters $(\bm{\omega}_\Theta,\bm{\omega}_C)$. However, as mentioned above, it is often complex to perform a joint sampling of the posterior distribution of all parameters in a copula model. This is due either to the dimensionality of the parameter vector or due to the structure of the associated complete conditional densities, which simultaneously involve both parameters of marginal distributions as well as copula parameters.

An alternative that we propose in this paper is to use the \textit{conditional approach} described in \cite{mirandaf2025}. This approach assumes the prior distribution $p(\bm{\omega}_\Theta,\bm{\omega}_C)$ has a structure like  $p(\bm{\omega}_\Theta)p(\bm{\omega}_C)$ or $p(\bm{\omega}_\Theta)p(\bm{\omega}_C|\bm{\omega}_\Theta)$, which is not unusual in the context of copula models. Thus, using that conditional approach, it is only required, on the one hand, to obtain the densities $p(\bm{\omega}_\Theta|\bm{D}_n)$ and $p(\bm{\omega}_C|\bm{\omega}_\Theta,\bm{D}_n)$, and on the other hand, to be able to sample from them.

We will consider the following prior specification
\[
p(\bm{\omega}_\Theta,\bm{\omega}_C)=p(\bm{\omega}_\Theta) p(\bm{\omega}_C).
\]
Thus, the posterior distribution is given up to a constant of proportionality by

\begin{equation*}
\begin{split}
&p(\bm{\omega}_\Theta,\bm{\omega}_C|\bm{D}_n) \\
&\propto \prod_{i=1}^n f(\bm{\theta}_i|\bm{\omega}_\Theta,\bm{\omega}_C)\times p(\bm{\omega}_\Theta,\bm{\omega}_C)\\
&=\prod_{i=1}^n c_w(F_1(\theta_{i1}|\bm{\omega}_1),\ldots,F_m(\theta_{im}|\bm{\omega}_m)|\bm{\omega}_C)\\
&\times \prod_{i=1}^n f_1(\theta_{i1}|\bm{\omega}_1)\cdots f_m(\theta_{im}|\bm{\omega}_m)\\
&\times p(\bm{\omega}_\Theta)p(\bm{\omega}_C).\\
\end{split}
\end{equation*}
In an equivalent way
\begin{equation}\label{eq7}
\begin{split}
p&(\bm{\omega}_\Theta,\bm{\omega}_C|\bm{D}_n)\\
&\propto \prod_{i=1}^n c_w(F_1(\theta_{i1}|\bm{\omega}_1),\ldots,F_m(\theta_{im}|\bm{\omega}_m)|\bm{\omega}_C)\times p(\bm{\omega}_C) \\
&\times \prod_{i=1}^n f_1(\theta_{i1}|\bm{\omega}_1)\cdots f_m(\theta_{im}|\bm{\omega}_m)\times p(\bm{\omega}_\Theta).
\end{split}
\end{equation}
From  Eq. (\ref{eq7}),  it is easy to show that
\begin{equation}\label{eq8}
\begin{split}
p(\bm{\omega}_\Theta|\bm{D}_n)\propto&\prod_{i=1}^n f(\theta_{i1}|\bm{\omega}_1)\cdots f(\theta_{im}|\bm{\omega}_m)\\
&\times p(\bm{\omega}_\Theta)
\end{split}
\end{equation}
and
\begin{equation}\label{eq9}
\begin{split}
p(&\bm{\omega}_C|\bm{\omega}_\Theta,\bm{D}_n)\\
\propto & \prod_{i=1}^n c_w(F_1(\theta_{i1}|\bm{\omega}_1),\ldots,F_m(\theta_{im}|\bm{\omega}_m)|\bm{\omega}_C)\\
&\times p(\bm{\omega}_C).
\end{split}
\end{equation}
Thus, we can use the densities (\ref{eq8}) and (\ref{eq9}) for sampling from the posterior distribution $p(\bm{\omega}_\Theta,\bm{\omega}_C|\bm{D}_n)$.

Next, the proposed methodology is specialized for two important copulas in the literature. However, the proposal discussed in this paper can be applied or generalized to other types of copulas.

\subsection{Two special cases}

In this section, we focus  the discussion above on the  Gaussian copula as well as the $t$ copula.
\subsubsection{The  $C_GPG$ model}

It  is said that a $m$-variate vector $\bm{\theta}=(\theta_1,\cdots,\theta_m)$, taking values in $\mathbb{H}$, has a Gaussian-coupled Projected Gamma model with parameters $\bm{\omega}_\Theta=(\bm{\omega}_1,\ldots,\bm{\omega}_m)$ and $\bm{\omega}_C=\bm{R}$, if its distribution function  is given by
\begin{equation*}
\begin{split}
F&(\theta_1,\ldots,\theta_m|\bm{\omega}_\Theta,\bm{\omega}_C)\\ 
&=C_G(F_{PG}(\theta_1|\bm{\omega}_1),\ldots,F_{PG}(\theta_m|\bm{\omega}_m)|\bm{\omega}_C)\\
&=\Phi_m(\mathcal{G}_1(\theta_1),\ldots,\mathcal{G}_m(\theta_m)| \bm{R}).
\end{split}
\end{equation*}
Here $\mathcal{G}_j(\theta_j)=\Phi^{-1}(F_{PG}(\theta_j|\bm{\omega}_j))$ and $\bm{\omega}_j=(\alpha_{j1},\alpha_{j2},\beta_j)$, where $\alpha_{j1},\alpha_{j2},\beta_j\in\mathbb{R}^+$  $\forall \ j=1,\ldots,m$ and $\bm{R}$ is a correlation matrix.  The corresponding density function is given by
\begin{equation*}
\begin{split}
C_GP&G(\theta_1,\ldots,\theta_m|\bm{\omega}_\Theta,\bm{\omega}_C) \\ 
=& \ c_G(F_{PG}(\theta_1|\bm{\omega}_1),\ldots,F_{PG}(\theta_m|\bm{\omega}_m)|\bm{\omega}_C)  \\
&\times PG(\theta_1|\bm{\omega}_1)\cdots PG(\theta_m|\bm{\omega}_m)\\
=&\dfrac{\phi_m(\mathcal{G}_1(\theta_1),\ldots,\mathcal{G}_m(\theta_m))\ |\bm{R})}{\phi(\mathcal{G}_1(\theta_1))\cdots \phi(\mathcal{G}_m(\theta_m))} \\
&\times PG(\theta_1|\bm{\omega}_1)\cdots PG(\theta_m|\bm{\omega}_m).
\end{split}
\end{equation*}
We propose the following prior specification
\begin{equation*}
\begin{split}
p(\bm{\omega}_\Theta,\bm{\omega}_C)&=p_\Theta(\bm{\omega}_\Theta)\times p_C(\bm{\omega}_C)\\
&=\prod_{j=1}^m p_j(\bm{\omega}_j)\times p_C(\bm{\omega}_C),
\end{split}
\end{equation*}
where
\begin{equation}\label{eq10}
p_j(\bm{\omega}_j)= \mathsf{Ga}(\alpha_{j1}|a_j,b_j) \mathsf{Ga}(\alpha_{j2}|c_j,d_j) \mathsf{Ga}(\beta_j|e_j,f_j)
\end{equation} 
and
\begin{equation}\label{eq11}
p_C(\bm{\omega}_C)=\prod_{1\leq r<m,\ r<q\leq m}\mathsf{Unif}(\rho_{rq}\ |\ \mbox{-}1,1),
\end{equation}
for $j=1,\ldots,m$. Here, $\rho_{rq}$ are the different correlation parameters of the matrix $\bm{R}$.

Thus, the posterior distribution is given up to a constant of proportionality by
\begin{equation}\label{posteriorG}
\begin{split}
&p(\bm{\omega}_\Theta,\bm{\omega}_C|\bm{D}_n)\\
\propto & \prod_{i=1}^n  \  c_G(F_{PG}(\theta_{i1}|\bm{\omega}_1),\dots , F_{PG}(\theta_{im}|\bm{\omega}_2)|\bm{R}) \\
\times & \  PG(\theta_{i1}|\bm{\omega}_1)\cdots PG(\theta_{im}|\bm{\omega}_m)  \\
\times & \prod_{j=1}^m \mathsf{Ga}(\alpha_{j1}|a_j,b_j)\cdot\mathsf{Ga}(\alpha_{j2}|c_j,d_j)\cdot\mathsf{Ga}(\beta_j|e_j,f_j)\\
\times & \prod_{1\leq r<m,\ j<q\leq m}  \mathsf{Unif}(\rho_{rq}\ |\ \mbox{-}1,1).
\end{split}
\end{equation}

\subsubsection{The  $C_TPG$ model}

It is said that a $m$-variate vector  $\bm{\theta}=(\theta_1,\cdots,\theta_m)$, taking values in $\mathbb{H}$, has a $t$-coupled Projected Gamma $t$-copula model  with parameters $\bm{\omega}_\Theta=(\bm{\omega}_1,\ldots,\bm{\omega}_m)$ and $\bm{\omega}_C=(\nu,\bm{R})$, if its distribution function  is given by
\begin{equation}
\begin{split}
F&(\theta_1,\ldots,\theta_m|\bm{\omega}_\Theta,\bm{\omega}_C)\\
&=C_T(F_{PG}(\theta_1|\bm{\omega}_1),\ldots,F_{PG}(\theta_m|\bm{\omega}_m)|\bm{\omega}_C)\\
&=T_m(\mathcal{T}_1(\theta_1),\ldots,\mathcal{T}_m(\theta_m)\ | \ \nu,\bm{R})
\end{split}
\end{equation}
Here $\mathcal{T}_j(\theta_j)=T^{-1}(F_{PG}(\theta_j|\bm{\omega}_j)|\nu)$ and $\bm{\omega}_j=(\alpha_{j1},\alpha_{j2},\beta_j)$, where $\alpha_{j1},\alpha_{j2},\beta_i\in\mathbb{R}^+$  $\forall \ j=1,\ldots,m$. For the matrix $\bm{R}$ to have an interpretation of correlation, we will take $\nu>2$.

The corresponding density function is given by
\begin{equation}
\begin{split}
C_TP&G(\theta_1,\ldots,\theta_m|\bm{\omega}_\Theta,\bm{\omega}_C)\\
=&c_T(F_{PG}(\theta_1|\bm{\omega}_1),\ldots,F_{PG}(\theta_m|\bm{\omega}_m)| \bm{\omega}_C) \\
&\times PG(\theta_m|\bm{\omega}_m)\cdots PG(\theta_m|\bm{\omega}_m)= \\
=&\dfrac{t_m(\mathcal{T}_1(\theta_1),\ldots,\mathcal{T}_m(\theta_m)|\nu,\bm{R})}{t(\mathcal{T}_1(\theta_1)|\nu)\cdots t(\mathcal{T}_m(\theta_m)|\nu)} \\
&\times PG(\theta_m|\bm{\omega}_m)\cdots PG(\theta_m|\bm{\omega}_m)
\end{split}
\end{equation}

In this case, we propose the following prior specification
\begin{equation}\label{eq13}
\begin{split}
&p(\omega_\Theta ,\bm{\omega}_C)\\
&= \prod_{j=1}^m \mathsf{Ga}(\alpha_{j1}|a_j,b_j)\cdot\mathsf{Ga}(\alpha_{j2}|c_j,d_j)\cdot\mathsf{Ga}(\beta_j|e_j,f_j)\\
&\times \prod_{1\leq r<m,\ r<q\leq m}\mathsf{Unif}(\rho_{rq}\ |\ \mbox{-}1,1)\times \mathsf{Ga}(\nu-2 | g,h).
\end{split}
\end{equation}
Thus, the posterior distribution is given up to a constant of proportionality by  
\begin{equation}\label{posteriorT}
\begin{split}
&p(\bm{\omega}_\Theta,\bm{\omega}_C|\bm{D}_n) \propto \\
 &\ \prod_{i=1}^n c_T(F_{PG}(\theta_{i1}|\bm{\omega}_1),\ldots,F_{PG}(\theta_{im}|\bm{\omega}_m)| \bm{\omega}_C)\\
&\times PG(\theta_m|\bm{\omega}_m)\cdots PG(\theta_m|\bm{\omega}_m)\\
 &\times \prod_{j=1}^m \mathsf{Ga}(\alpha_{j1}|a_i,b_i)\mathsf{Ga}(\alpha_{j2}|c_j,d_j)\mathsf{Ga}(\beta_j|e_j,f_j)\\
  &\times  \prod_{1\leq r<m,\ r<q\leq m}\mathsf{Unif}(\rho_{rq}\ |\ \mbox{-}1,1)\times \mathsf{Ga}(\nu-2 | g,h).
\end{split}
\end{equation}

\section{Illustrations}\label{sec4}

In this section, we present three examples to illustrate the performance of the  proposed methodology. In the first example, we analyze a simulated data set with a 2-variate distribution $C_GPG$, while in the second, the data follow a 3-variate distribution $C_TPG$. A third example with real data is presented to show the application of this type of multivariate models for circular bounded data in an orthopedic context. In all of these examples, we impose non-informative priors by setting  $a_s=c_s=e_s=1$ and $b_v=d_v=f_v=0.2$, for all necessary subscripts $s$ and $v$.

\subsection{Example 1}

In this example, we examine data simulated from a distribution $C_GPG$. Specifically, we consider a data sample of angles $(\theta_1,\theta_2)$ of size $n=500$ from the following model:
\begin{equation}
\begin{split}
C_GPG&(\theta_1,\theta_2|\bm{\omega}_\Theta,\bm{\omega}_C)\\
=&\  c_G(F_{PG}(\theta_1|\bm{\omega}_1),F_{PG}(\theta_2|\bm{\omega}_2)|\bm{R})\\
&\times PG(\theta_1|\bm{\omega}_1)PG(\theta_2|\bm{\omega}_2)\\
\end{split}
\end{equation}
where 
\begin{equation*}
\bm\omega_1=(\alpha_{11},\alpha_{12},\beta_1)=(2,2,1),
\end{equation*}
\begin{equation*}
\bm\omega_2=(\alpha_{21},\alpha_{22},\beta_2)=(0.5,0.5,1),
\end{equation*}
and
\begin{equation*}
\bm{R}=\begin{bmatrix}
1 & \rho_{12}\\
\rho_{12} & 1
\end{bmatrix}=\begin{bmatrix}
1 & 0.7\\
0.7 & 1
\end{bmatrix}.
\end{equation*}

\begin{figure}[h]
\begin{center}
\includegraphics[scale=0.5]{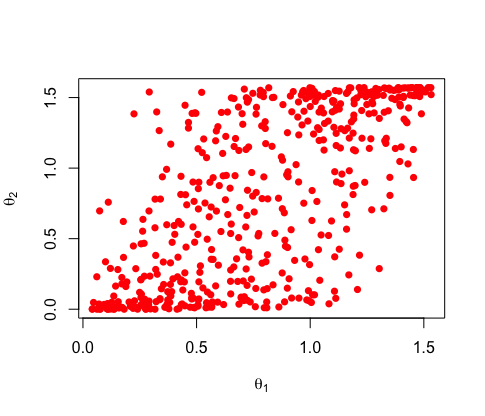}
\end{center}
\caption{Cartesian representation of the data of Example 1.}
\label{Fig1}
\end{figure}

\begin{figure}[h]
\begin{center}
\includegraphics[scale=0.4]{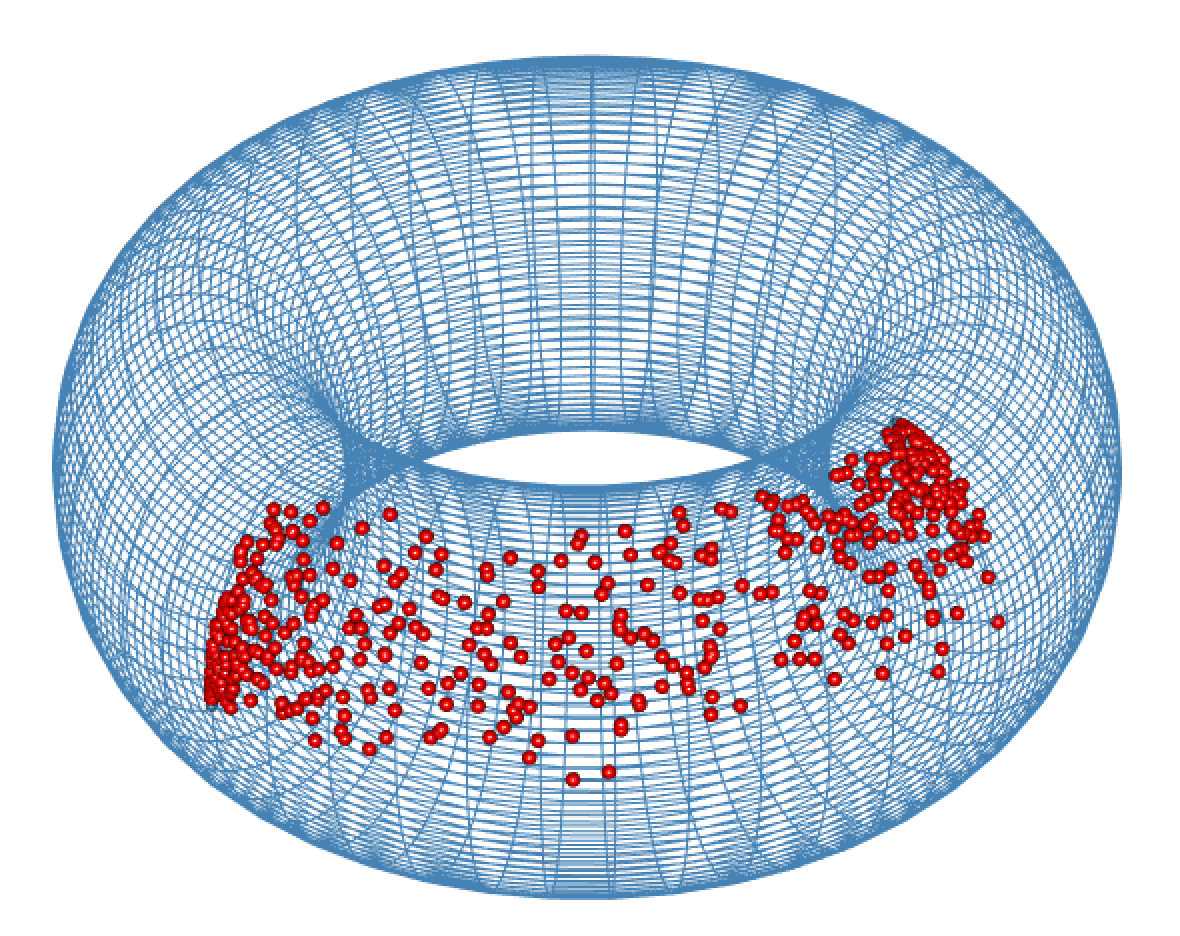}
\end{center}
\caption{Representation on the unit torus for data of Example 1.}
\label{Fig2}
\end{figure}

Figure \ref{Fig1} shows a Cartesian representation of that data set. On the other hand, Figure \ref{Fig2} shows the data on the surface of the unit torus. It can be noted this distribution exhibits a certain symmetry.

We use the conditional approach to simulate a sample of size 1000 from the posterior distribution  (\ref*{posteriorG}).  For the first stage, a Gibbs sampling was used to sample from the posterior distribution $p(\bm{\omega}_1, \bm{\omega}_2 |\bm{D}_n)$. For the second step, we used an  {\it adapting rejection metropolis sampling} algorithm (arms) to sample from the posterior distribution $p(\rho_{12}| \bm{\omega}_1,\bm{\omega}_2,\bm{D}_n)$. Convergence diagnostics led us to stop the simulating process after 120,000 iterations, discarding the first 70,000 as burn-in. From the remaining 50,000 iterations, we kept one observation every 50 iterations as part of the final sample.

\begin{figure*}
\begin{center}
\includegraphics[scale=0.3]{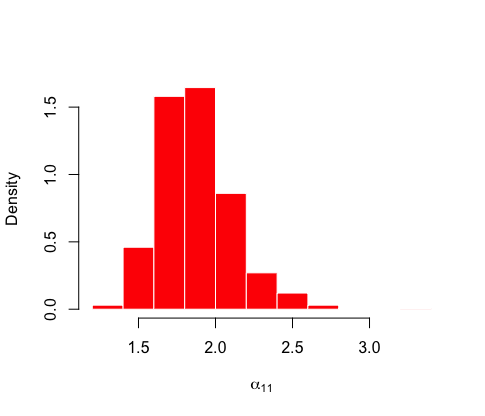}
\includegraphics[scale=0.3]{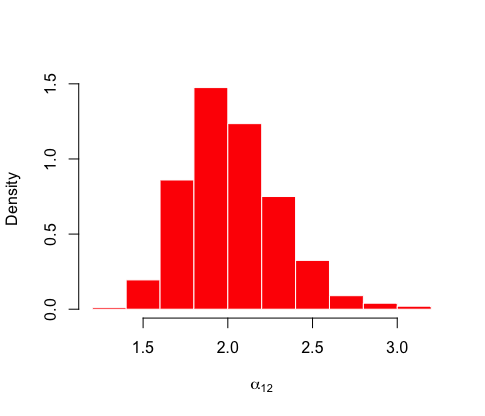}
\includegraphics[scale=0.3]{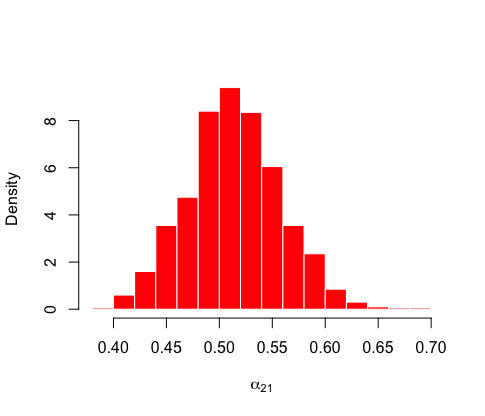}
\includegraphics[scale=0.3]{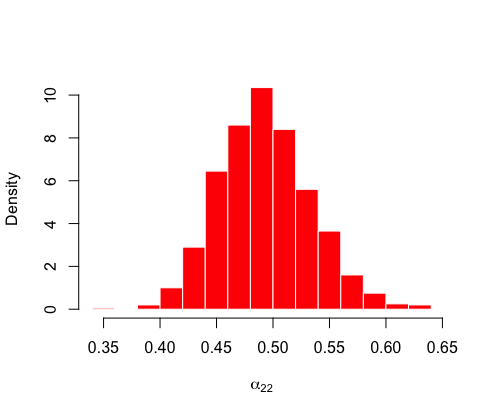}
\includegraphics[scale=0.3]{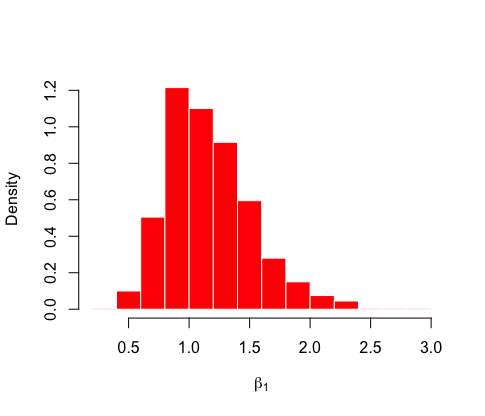}
\includegraphics[scale=0.3]{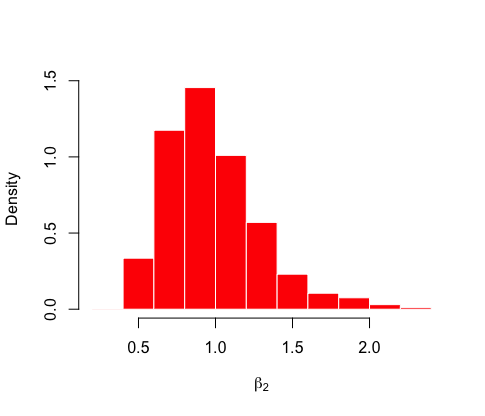}
\includegraphics[scale=0.3]{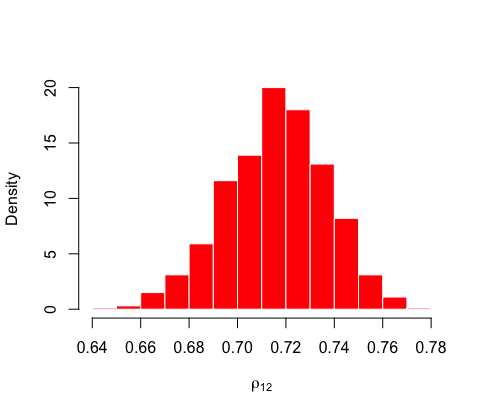}
\end{center}
\caption{Component-wise posterior distributions of the parameter vectors $\bm{\omega}_1$ and $\bm{\omega}_2$ and the parameter $\rho_{12}$.}
\label{Fig3}
\end{figure*}

Figure  \ref{Fig3} shows the component-wise marginal distributions for the vectors  $\bm{\omega}_1$ and  $\bm{\omega}_1$, as well as the parameter $\rho_{12}$. On the other hand, Table \ref{tabla1} presents the $95\%$ posterior credible intervals for each elements of $\bm{\omega}_1$ and  $\bm{\omega}_2$, and  $\rho_{12}$. It can be seen the true value of each of the parameters is well inside the corresponding credible interval.

\begin{table}[h]
	\centering
	\caption{95\% credible intervals for each parameters of vectors $\bm\omega_1$, $\bm\omega_2$  and $\rho_{12}$.}
	\begin{tabular}{ccc}
		\hline 
		\text{Parameter}&\text{Actual value}& \text{Credible interval} \\ \hline
		$\alpha_{11}$ & 2 & ( 1.45, 2.38 ) \\
		$\alpha_{12}$ & 2 & ( 1.54, 2.57 )\\
		$\beta_1$ & 1 & ( 0.55, 1.89 ) \\
		$\alpha_{21}$ & 0.5 & ( 0.42, 0.59 ) \\
		$\alpha_{22}$ & 0.5 & ( 0.42, 0.58 )\\
		$\beta_2$ & 1 & ( 0.43, 1.59 ) \\
		$\rho_{12}$ & 0.7 & ( 0.68, 0.75 ) \\
		\hline
	\end{tabular}
	\label{tabla1}
\end{table}

\subsection{Example 2}
For this example, a random sample $(\theta_{1i},\theta_{2i},\theta_{3i})$ of size 500 was simulated. The sample was generated from the following  $C_TPG$ model.
\begin{equation}
\begin{split}
C_TPG&(\theta_1,\theta_2,\theta_3|\bm{\omega}_\Theta,\bm{\omega}_C)\\
=&\  c_G(\ F_{PG}(\theta_1|\bm{\omega}_1),F_{PG}(\theta_2|\bm{\omega}_2), F_{PG}(\theta_3|\bm{\omega}_3 )\ |\ \bm{R}, \nu \ )\\
&\times PG(\theta_1|\bm{\omega}_1) PG(\theta_2|\bm{\omega}_2) PG(\theta_3|\bm{\omega}_3)\\
\end{split}
\end{equation}
where
\begin{equation*}
\bm\omega_1=(\alpha_{11},\alpha_{12},\beta_1)=(2,2,2),
\end{equation*}
\begin{equation*}
\bm\omega_2=(\alpha_{21},\alpha_{22},\beta_2)=(0.5,3,1),
\end{equation*}
\begin{equation*}
\bm\omega_3=(\alpha_{31},\alpha_{32},\beta_3)=(3,5,3),
\end{equation*}
\begin{equation*}
\nu=3,
\end{equation*}
and
\begin{equation*}
\bm{R}=\begin{bmatrix}
1 & \rho_{12} & \rho_{13} \\
\rho_{12} & 1 & \rho_{23} \\
\rho_{13} & \rho_{23} & 1 \\
\end{bmatrix}=\begin{bmatrix}
1 & 0.75 & -0.75 \\
0.75 & 1 & -0.75 \\
-0.75 & -0.75 & 1 \\
\end{bmatrix}.
\end{equation*}

\begin{figure}[H]
\begin{center}
\includegraphics[scale=0.5]{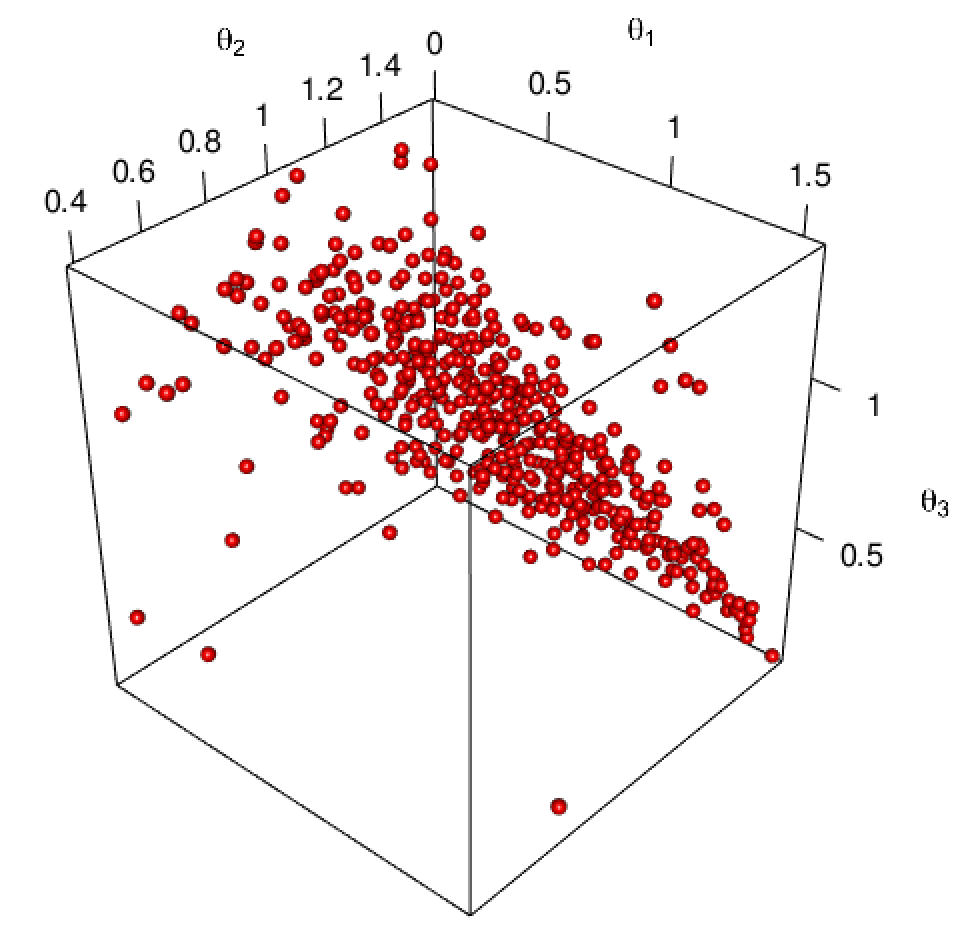}
\end{center}
\caption{Cartesian representation of the  data of Example 2.}
\label{Fig4}
\end{figure}

Figure \ref{Fig4} displays the data in a three-dimension Cartesian representation. It can be noted this configuration produces a data set with some asymmetry.

We use our conditional approach to simulate a sample of size 1000 from the posterior distribution  (\ref*{posteriorT}).  For the first stage, a Gibbs sampling was used to sample the posterior distribution $p(\bm{\omega}_1, \bm{\omega}_2, \bm{\omega}_3  |\bm{D}_n)$. For the second step, we use again  a Gibbs sample to sample from the posterior distribution $p(\bm{R},\nu |\bm{D_n})$. The latteris based on the complete conditional densities $p(\rho_{rs}| \bm{\omega}_1, \bm{\omega}_2,  \bm{\omega}_3,\nu, \bm{D}_n)$ $\forall \ 1 \le r <s \le 3$,  and $p(\nu|\bm{\omega}_1,\bm{\omega}_2,\bm{\omega}_3,\{\rho_{rs}\}, \bm{D}_n)$. Convergence diagnostics led us to stop the simulating process after $100,000$ iterations, discarding the first $50,000$ as burn-in. From the remaining $50,000$ iterations, we kept one observation every $50$ iterations as part of the final sample.

\begin{figure*}[tp]
\begin{center}
\includegraphics[scale=0.275]{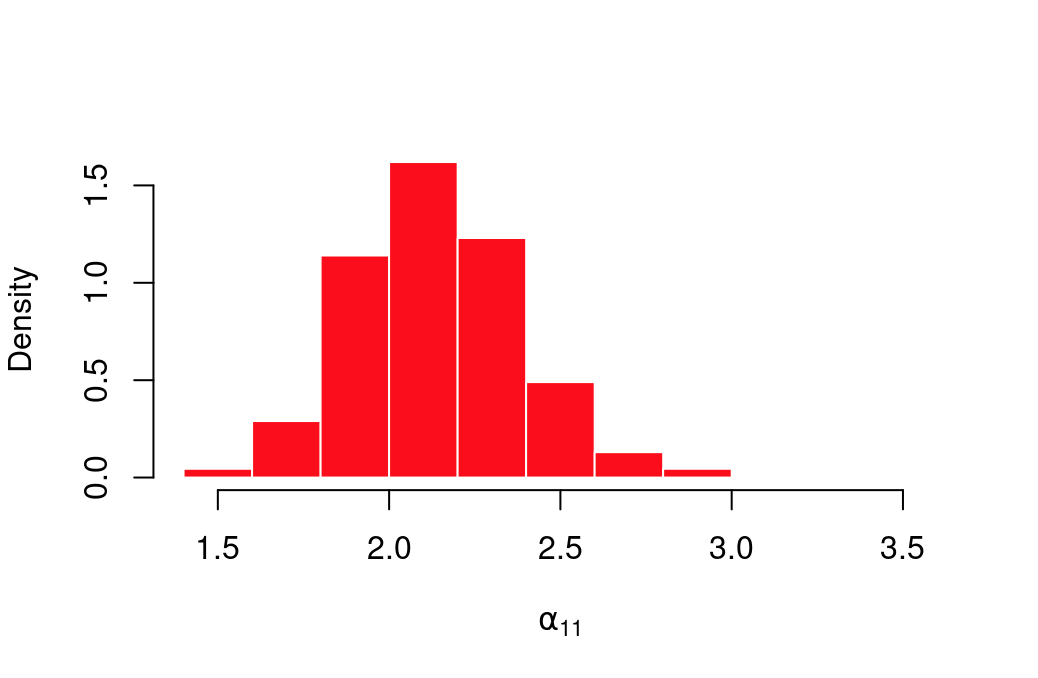}
\includegraphics[scale=0.275]{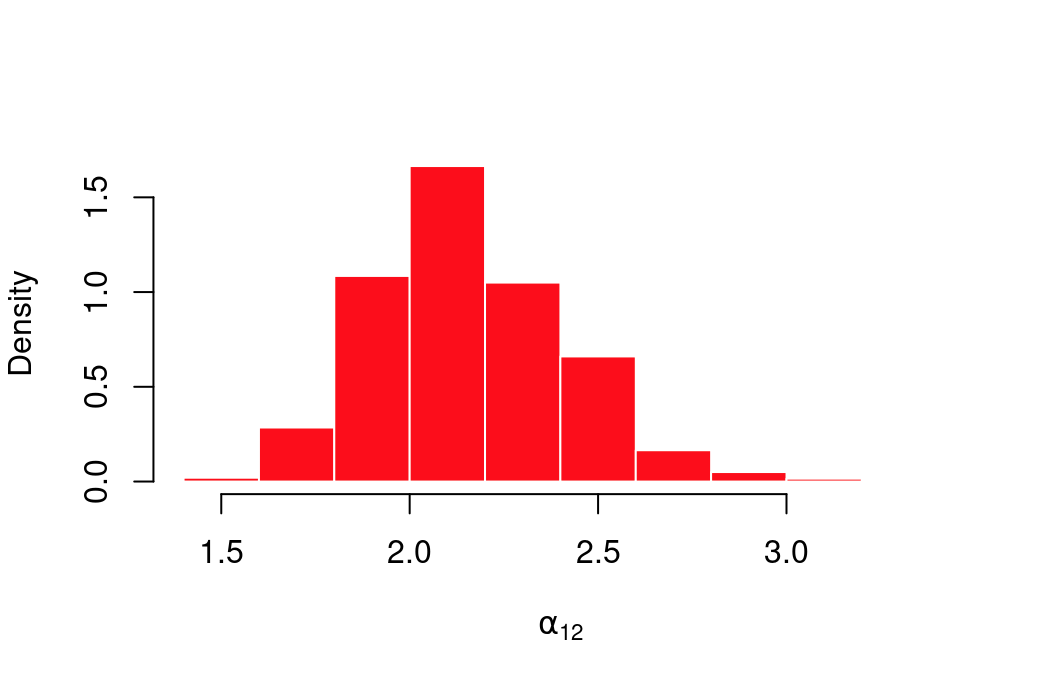}
\includegraphics[scale=0.275]{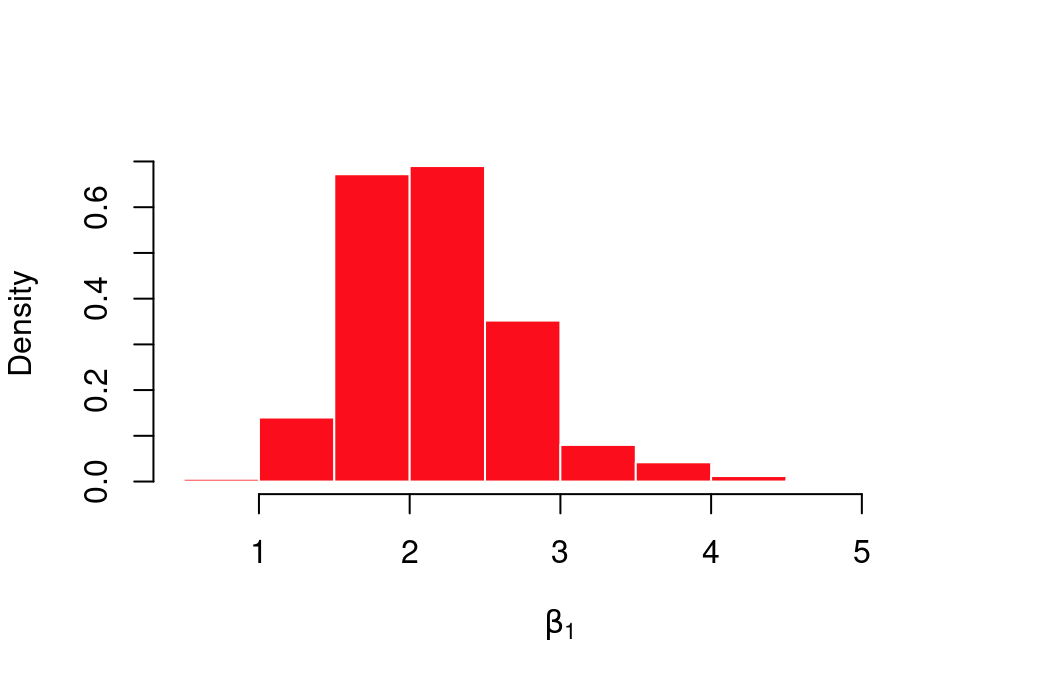}
\includegraphics[scale=0.275]{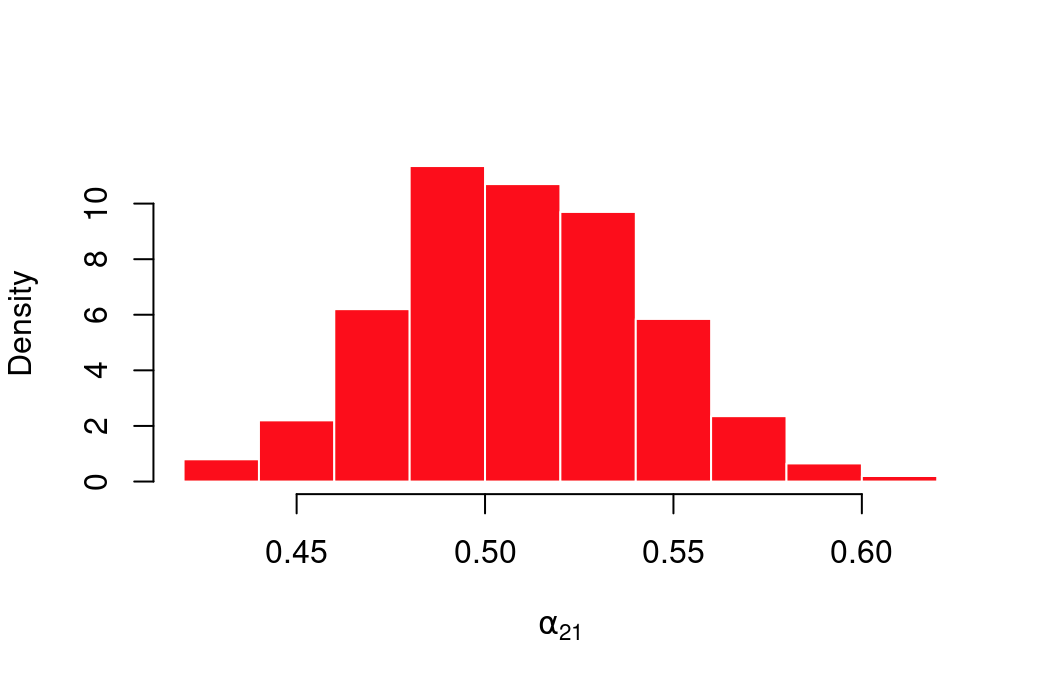}
\includegraphics[scale=0.275]{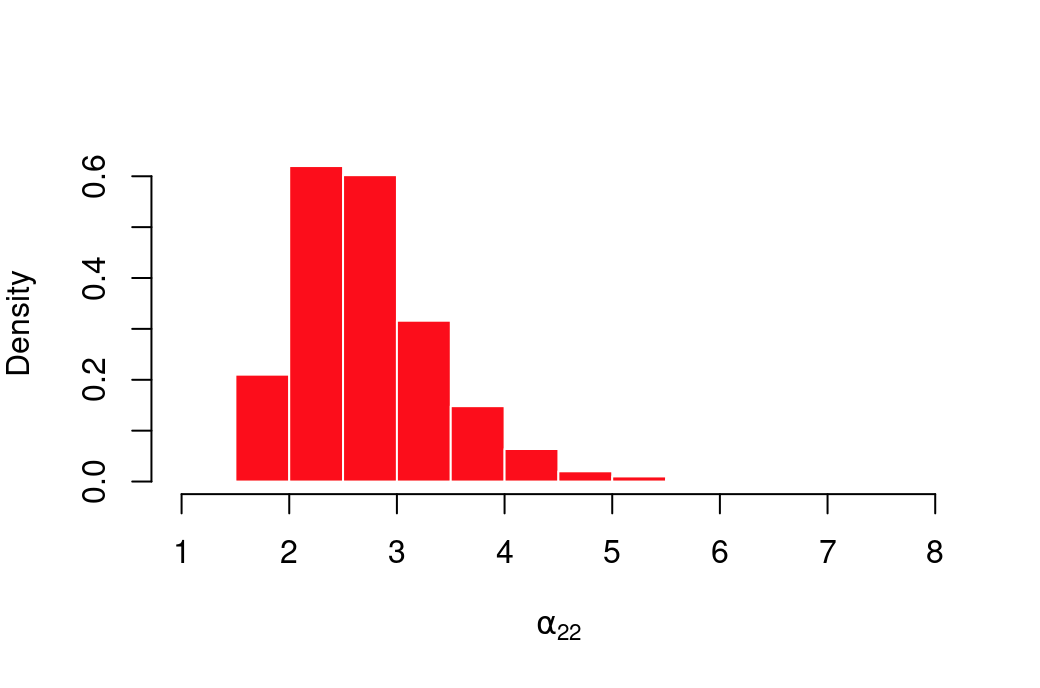}
\includegraphics[scale=0.275]{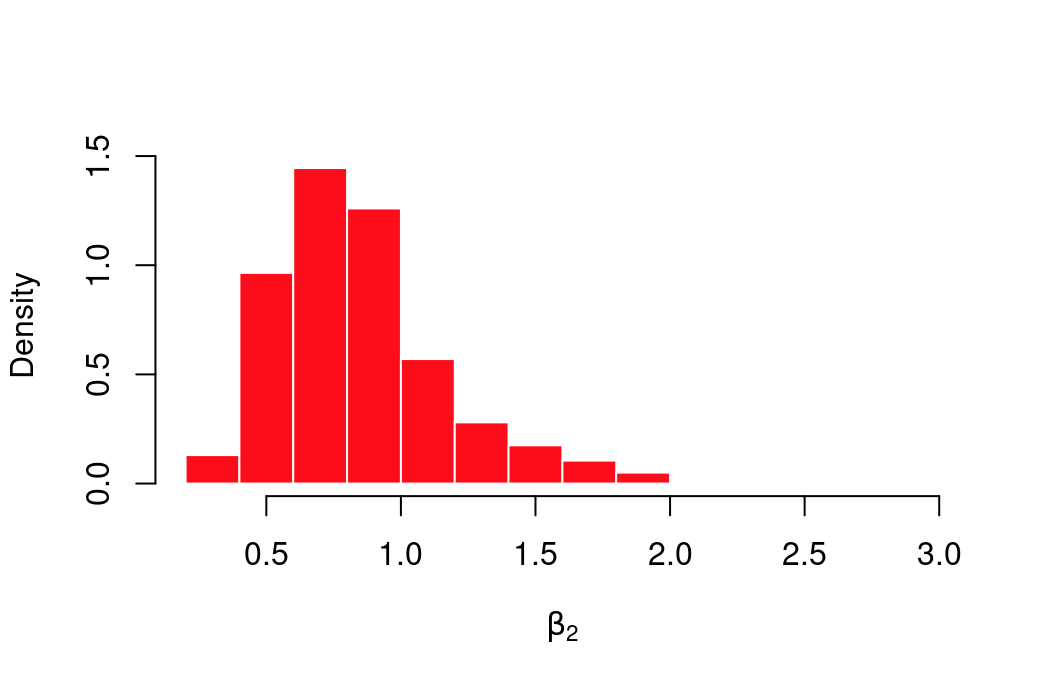}
\includegraphics[scale=0.275]{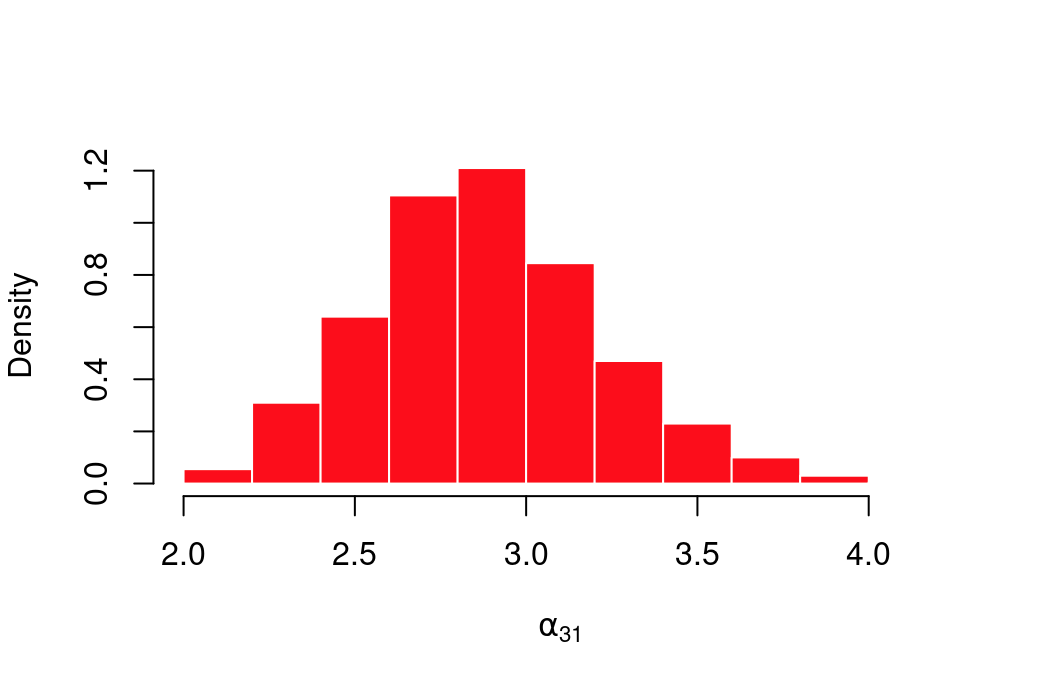}
\includegraphics[scale=0.275]{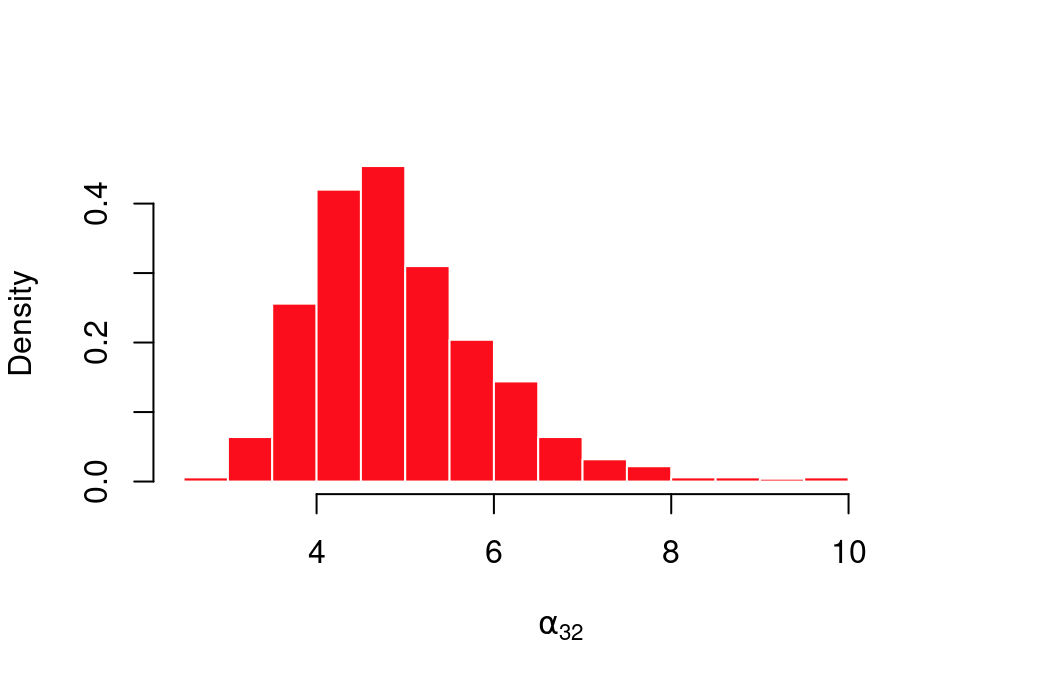}
\includegraphics[scale=0.275]{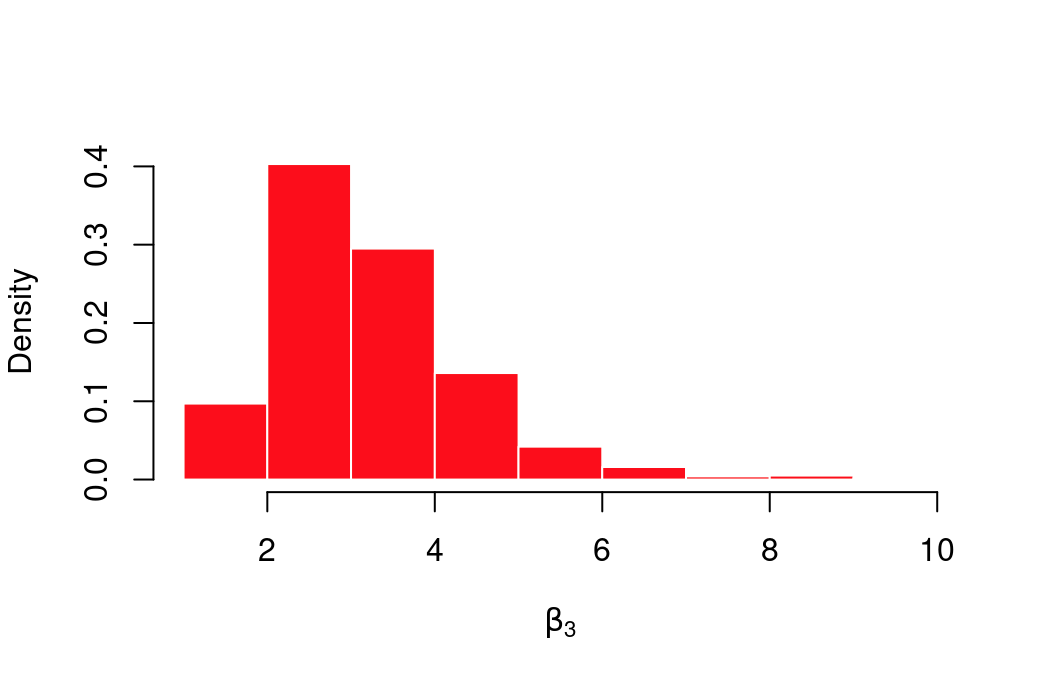}
\end{center}
\caption{Posterior distributions for component-wise of parameter vectors $\bm{\omega}_1$ and $\bm{\omega}_2$ of Example 2.}
\label{Fig5}
\end{figure*}

\begin{figure*}
\begin{center}
\includegraphics[scale=0.35]{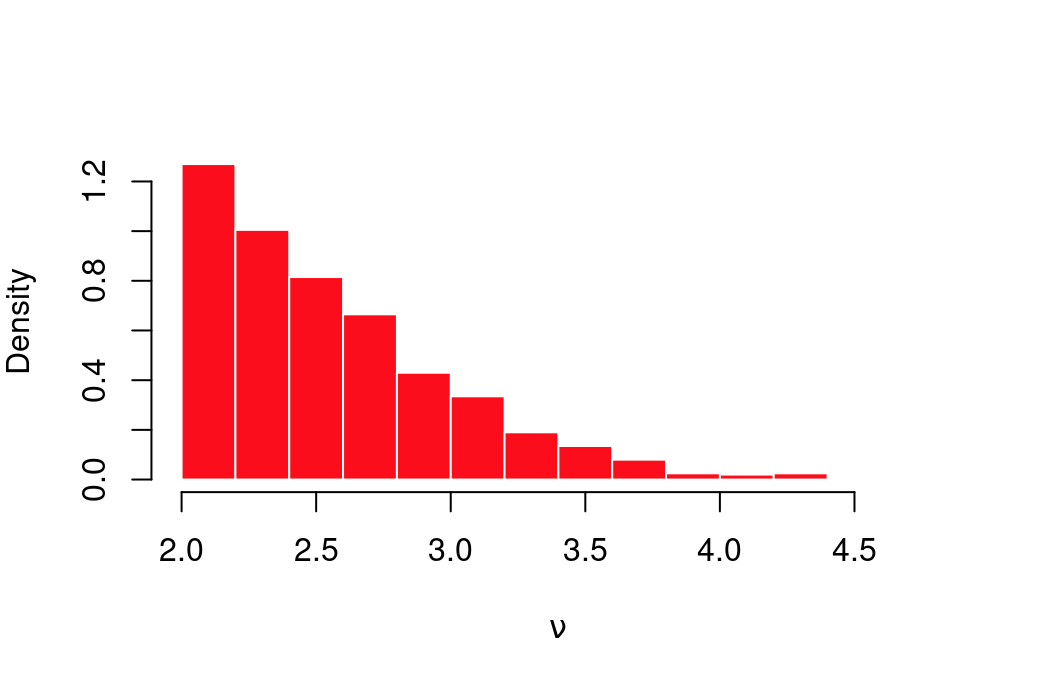}
\includegraphics[scale=0.35]{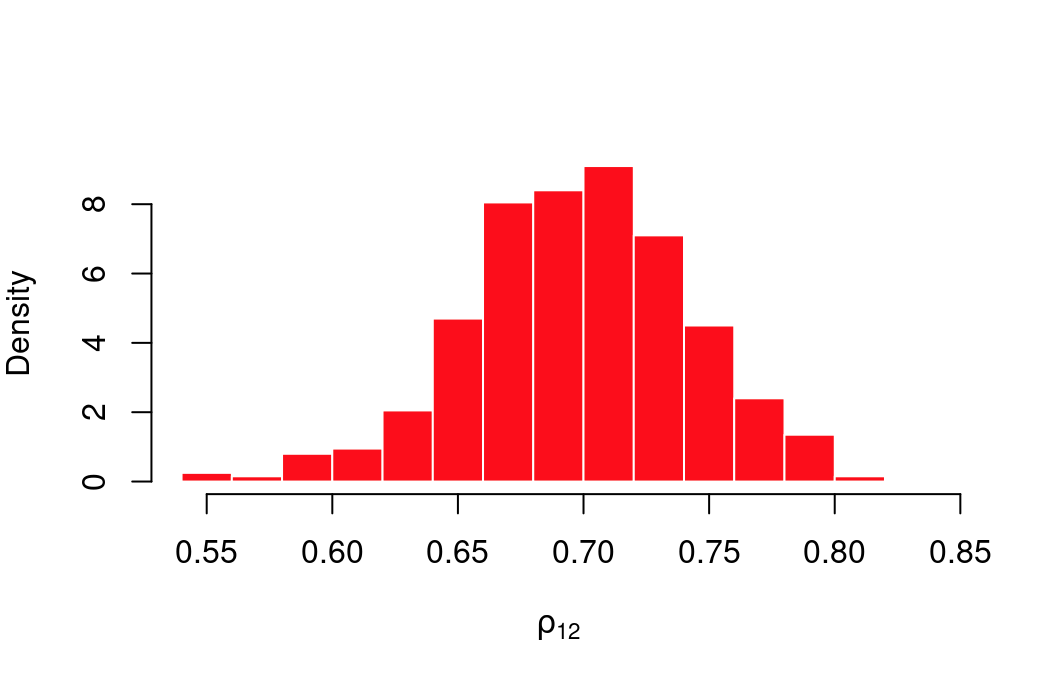}
\includegraphics[scale=0.35]{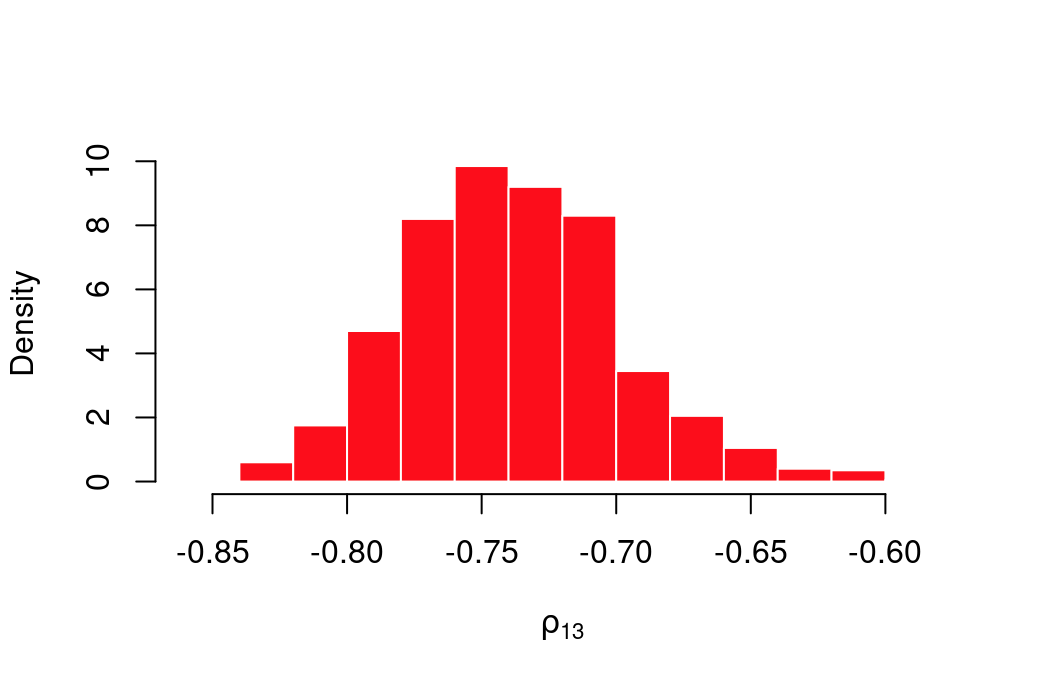}
\includegraphics[scale=0.35]{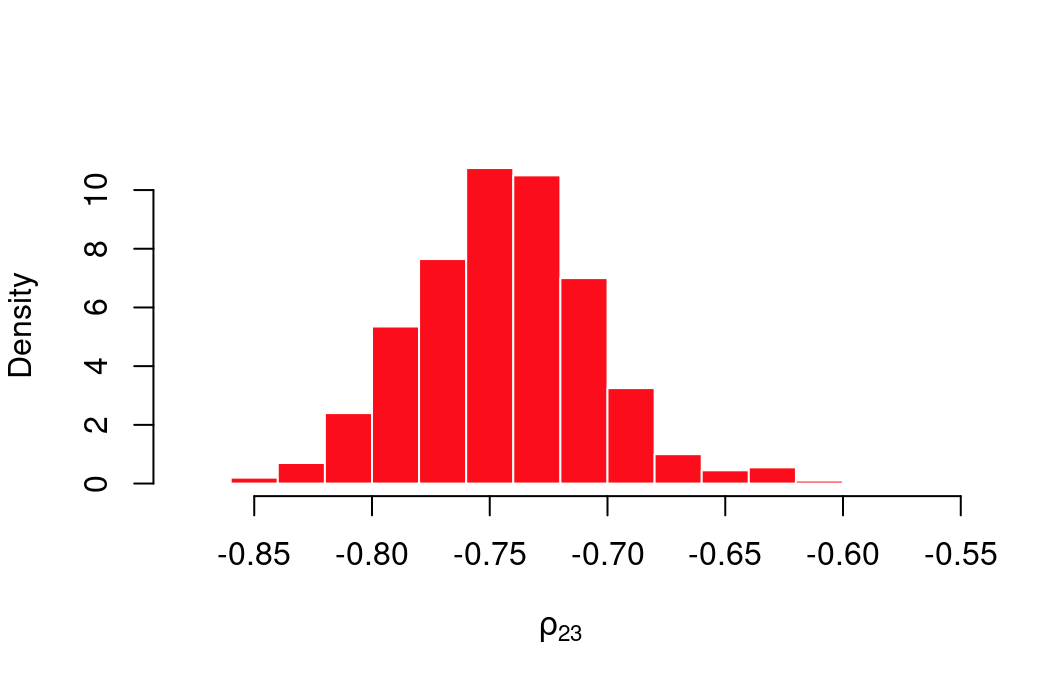}
\end{center}
\caption{Posterior distributions for component-wise of copula parameters  $\nu$ and $\bm{R}$ of Example 2.}
\label{Fig6}
\end{figure*}

Figure \ref{Fig5} presents the component-wise marginal distributions for the vectors $ \bm{\omega}_1$, $ \bm{\omega}_2$ and $ \bm{\omega}_3$.  On the other hand, Figure \ref{Fig6} shows the posterior distributions for the copula parameters, $\nu,\rho_{12},\rho_{13}$ and $\rho_{23}$. Finally, Table \ref{tabla2}  presents the $95\%$ posterior credible intervals for each elements of $\bm{\omega}_1$,  $\bm{\omega}_2$ and $\bm{\omega}_3$, as well as for the parameters  $\nu,\rho_{12},\rho_{13}$ and $\rho_{23}$. It can be seen the true value of each of the parameters is well inside the corresponding credible interval.

\begin{table}
	\centering
		\caption{95\% credible intervals for each parameters of vectors $\bm{\omega}_1$, $\bm{\omega}_2$,  $\bm{\omega}_3$ and $\bm{\omega}_C$ of Example 2.}
	\begin{tabular}{ccc}
		\hline 
		\text{Parameter}&\text{Actual value}& \text{Credible interval} \\ \hline
		$\alpha_{11}$ & 2 & ( 1.65, 2.60 ) \\
		$\alpha_{12}$ & 2 & ( 1.67, 2.66 )\\
		$\beta_1$ & 2 & ( 1.11, 3.27 ) \\
		$\alpha_{21}$ & 0.5 & ( 0.45, 0.57 ) \\
		$\alpha_{22}$ & 3 & ( 1.50, 4.03 )\\
		$\beta_2$ & 1 & ( 0.34, 1.53 ) \\
		$\alpha_{31}$ & 3 & ( 2.26, 3.57 ) \\
		$\alpha_{32}$ & 5 & ( 3.25, 7.00 )\\
		$\beta_3$ & 3 & ( 1.46, 5.41 ) \\
		$\nu$ & 3 & ( 2.00, 3.44 ) \\
		$\rho_{12}$ & 0.75 & ( 0.62, 0.80 ) \\
		$\rho_{13}$ & -0.75 & ( -0.82, -0.66 ) \\
		$\rho_{23}$ & -0.75 & (-0.82, -0.67 ) \\
		\hline
	\end{tabular}
	\label{tabla2}
\end{table}

\subsection{Real data example}

Gait assessments are important in biomechanical and movement studies as they identify biomechanical situations that could lead to future musculoskeletal problems in patients. Abnormal gait mechanics are particularly important for recognizing musculoskeletal pathologies that specifically affect the lower limbs (see, for example, \cite{higginson2008} and \cite{harradine2018}). An abnormal gait is often accompanied by an irregular foot angle (see, for example, \cite{rawlings2023}). Thus, the study of specific angles between both feet is of interest in gait analysis. In particular, the study of angles such as the Fick angle, which is a measure between the midline of the body and the degree of external rotation, is important  in gait studies. However, there are few researchers analyzing this association (\cite{rawlings2023}).

\begin{figure}[h]
\begin{center}
\includegraphics[scale=0.5]{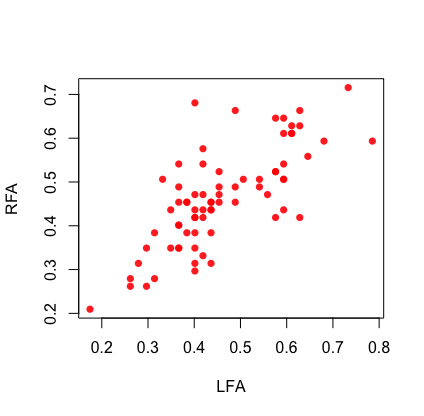}
\end{center}
\caption{Sample of Fick angles $(\theta_L,\theta_R)$ corresponding to the right foot (RFA) and the left foot (LFA) from 73 orthopedic patients.}
\label{Fig7}
\end{figure}

For this example we considered data from 73 orthopedic patients. For each patient, the Fick angle from both the left foot (LFA) and the right foot (RFA) were measured. Data can be obtained from the Mendeley Data Repository through the URL \url{https://data.mendeley.com/datasets/f9gs9rr2ng}, see also \cite{rawlings2023}. It must be noted that by nature, each of these  angles is only defined within the arc $(0,\pi /2]$. A Cartesian scatter plot of these angles data $\{(\theta_{L,1},\theta_{R,1}),\dots, (\theta_{L,73},\theta_{R,73})\}$ is shown in Figure \ref{Fig7}.

We fit a $C_GPG$ model as well as a $C_TPG$ model to describe the joint behavior of both angles $(\theta_L,\theta_R)$. For each of  the models, a sample of size 1,000 from the corresponding posterior distribution was obtained. For the $C_GPG$ model, a burn-in period of 100,000 iterations was established and a lag of 50 iterations between each element for the final sample was considered. For the $C_TPG$ model, a burn-in period of 250,000 iterations was necessary and a lag of 50 iterations between each element for the final sample was established.

\begin{table}[tp]
	\centering
		\caption{95\% credible intervals for component-wise of parameter vectors $\bm{\omega}_1$, $\bm{\omega}_2$  and $\bm{\omega}_C$ for the  $C_GPG$ and $C_TPG$ models.}
	\begin{tabular}{c c c} \hline
		\multirow{2}{20mm}{Parameter} &
		\multicolumn{2}{c}{Credible Intervals} \\  
		& Model $C_GPG$ & Model $C_TPG$ \\
		\hline
		$\alpha_{11}$ & (9.267, 33.632) & (8.862, 29.942) \\
		$\alpha_{12}$ & (9.791, 37.490) & (10.330, 41.036)\\
		$\beta_1$ &  ( 0.472, 6.246) & ( 0.525,  6.964) \\
		$\alpha_{21}$  & (9.160, 38.040)& (9.607, 35.163) \\
		$\alpha_{22}$  & (10.778, 39.277) & (10.614, 40.346)\\
		$\beta_2$ &  (0.477, 5.762)&  (0.551,  6.067) \\
		$\rho$ & (0.654, 0.844) & (0.622,  0.862) \\
		$\nu$ & --& (2.003, 22.893) \\
		\hline
	\end{tabular}
	\label{tabla3}
\end{table}

\begin{figure}[tp]
	\begin{center}
		\includegraphics[scale=0.45]{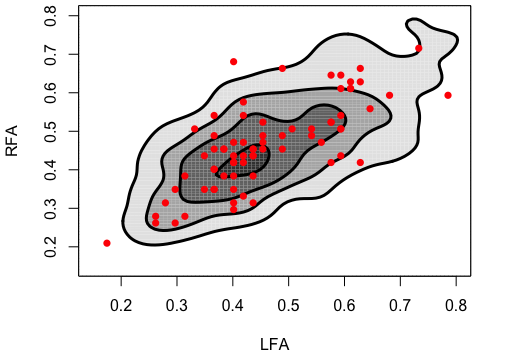}
		\includegraphics[scale=0.45]{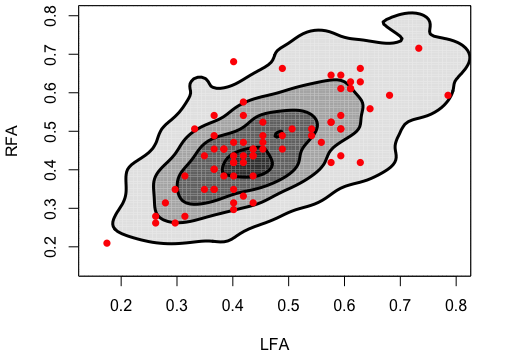}
	\end{center}
	\caption{The upper figure shows the level curves of predictive distribution for the $C_TPG$ model. The level curves of the predictive distribution for the $C_GPG$ model are shown in the lower figure.}
	\label{Fig8}
\end{figure}

Table \ref{tabla3} shows the 95\% posterior credible intervals for each parameter of the $C_GPG$ and $C_TPG$ models. On the other hand, the final predictive distributions for both models are presented in Figure \ref{Fig8}. Initially, both models seem competitive, so a way to compare them is necessary. We consider the predictive measure LPML (the logarithm of the pseudo marginal likelihood) originally suggested by \cite{geissereddy1979}. For the  $C_GPG$ model, a LPML value of -59.8 was obtained, while for the $C_TPG$ model, a value of -59.5 was obtained. From a predictive point of view, these results support the hypothesis that the model obtained with a t-copula is slightly better than the one obtained with a Gaussian-copula.

The results shown in this section suggest  the methodology proposed in this work is suitable for defining and analyzing multivariate models whose marginal densities are defined by circular distributions bounded on the arc $(0, 2\pi]$.

\section{Concluding remarks}\label{sec5}

This work introduces a methodology for constructing multivariate models whose marginal variables are defined only in the first quadrant of the unit circle. The proposed methodology is based on the construction of multivariate models defined by copula functions where the marginal densities are Projected Gamma distributions. Although the idea of building multivariate distributions using copulas is not new, when working with marginal densities that are both directional and bounded, the problem becomes challenging. In addition, even if there are different strategies to make Bayesian inferences in copula models, in this work we exploit a conditional approach that takes advantage of the structure of a copula model  to make inferences in a more efficient way. \\

In this paper we focus on implicit copulas, such as the Gaussian copula and the $t$ copula. However, with some adjustments, the proposed methodology can be adapted to consider other types of copulas. Finally, it is worth mentioning that this approach to build multivariate models for bounded directional data can be enriched by considering more general marginal distributions such as those derived from nonparametric contexts. These generalizations are currently under development.

\bmhead{Acknowledgements}The work of the first author was supported by CONAHCYT, Mexico. The second author was partially supported by CONAHCYT, through the Sistema Nacional de Investigadores e Investigadoras, Mexico. The support received from the Department of Mathematics of the Metropolitan Autonomous University, Iztapalapa Unit is also gratefully acknowledged.

\bibliography{Montesinos_Nunez}

\end{document}